\documentclass{cup-hpl}

\begin{document}
\bibliographystyle{unsrtnat}




\title{Comparative study of second harmonic generation at 1030~nm in BiBO and LBO crystals using a 100~W-class picosecond laser}

\author[1,2]{Huzefa Aliasger}
\author[1,3]  {Šimon Šatra\corresp{Huzefa Aliasger, HiLASE Centre, FZU - Institute of Physics of the Czech Academy of Sciences, Za Radnicí 828, 25241 Dolní Břežany, Czechia.   \email{huzefa.bhanpurwala@hilase.cz}}}
\author[1]{Ondřej Novák}
\author[1]{Jiří Mužík}
\author[2]{Michal Jelínek}
\author[1]{Martin Smrž}
\author[1]{Tomáš Mocek}

\address[1]{HiLASE Centre, FZU - Institute of Physics of the Czech Academy of Sciences, Dolní Břežany, Czech Republic}

\address[2]{
	Department of Laser Physics and Photonics, Faculty of Nuclear Sciences and Physical Engineering, Czech Technical University in Prague, Prague, Czech Republic
}

\address[3]{Faculty of Mathematics and Physics, Charles University, Prague, Czech Republic}

\begin{abstract}
	
\noindent We present a systematic experimental comparison of single-pass second-harmonic generation (SHG) in bismuth triborate (BiBO) and lithium triborate (LBO) nonlinear crystals, driven by a 1.3~ps, 91~kHz laser at 1030~nm with up to 57~W of average input power. Both crystals yielded 32~W of second harmonic (SH) output at 515~nm, corresponding to a conversion efficiency of 56\%, which to the best of our knowledge represents the highest SH output power reported in the green spectral region using a BiBO crystal. Power dependence, long-term stability, beam quality, pulse duration, spectral properties, thermal effects, and angular acceptance bandwidth are characterized and directly compared for both crystals. These results provide quantitative performance benchmarks to guide the selection of nonlinear crystals for high-average-power, ultrashort-pulse frequency conversion near 1030~nm.

\end{abstract}


\maketitle

\section{Introduction}

Second harmonic generation (SHG) is one of the most efficient ways of extending the spectral coverage of solid-state lasers. In particular, SHG in the green spectral region is of considerable practical importance for a range of applications including laser micromachining \cite{ma13132962}, pumping of optical parametric oscillators and amplifiers \cite{Sukeert:19, Galletti_Oliveira}, pumping of titanium:sapphire lasers \cite{Eichner_25},  scientific instrumentation \cite{s23010292}, and medical diagnostics \cite{PARK2020100204}. The rapid progress in high-average-power picosecond and sub-picosecond thin-disk \cite{Xu_Gao_2024, Osolodkov_26}, fiber \cite{Fathi2025}, InnoSlab \cite{6843864}, and zig-zag slab \cite{Yang_26} laser systems has significantly increased the demand for nonlinear frequency-conversion stages that can simultaneously provide high conversion efficiency, excellent beam quality, and long-term operational stability, particularly for operation in the near-infrared spectral region around 1030~nm.

Lithium triborate (LBO, $\mathrm{LiB_3O_5}$) has long been the workhorse nonlinear crystal for high-power SHG in this wavelength range, owing to its favorable combination of relatively high damage threshold (2.5~GW/cm² at 1064 nm, 10 ns \cite{Raicol_LBO_2026}), low walk-off, moderate effective nonlinearity, and broad transparency. These properties have made LBO a standard choice in large-scale and industrial laser systems. In contrast, bismuth triborate (BiBO, $\mathrm{BiB_3O_6}$) has attracted growing interest as an alternative frequency-doubling crystal, offering a substantially larger effective nonlinear coefficient, which is more than four times that of LBO and in principle enables efficient SHG in significantly shorter crystal lengths. The reduced crystal length can mitigate deleterious effects such as group velocity dispersion (GVD) and temporal walk-off, which become increasingly relevant in the picosecond regime. However, BiBO exhibits a considerably lower damage threshold (0.3~GW/cm² at 1064~nm, 10~ns \cite{Crysmit_BiBO_2026}), larger spatial walk-off, and typically higher absorption at the second harmonic (SH) wavelength, which may enhance thermal loading and affect long-term phase-matching stability under high-power operation.

While numerous studies have independently characterized SHG performance in the LBO crystal for high-average-power lasers of over 50~W \cite{XU2025111963, MA2025112567}, the BiBO crystal has not been tested for such high-average-power in picosecond regime. Systematic experimental studies that directly compare nonlinear optical crystals under identical high-average-power, picosecond driving conditions remain notably scarce in the literature. Such direct comparisons are indispensable for rigorously quantifying the practical trade-offs governing nonlinear conversion efficiency, angular acceptance bandwidth, thermal phase-matching robustness, beam quality preservation, and output power stability, all of which are parameters of critical importance in the design and optimization of next-generation harmonic generation modules integrated with high-average-power laser platforms.

In this work, we present a comprehensive experimental comparison of single-pass SHG in BiBO and LBO crystals driven by a 1.3~ps, 91~kHz, 57~W Yb:YAG thin-disk laser at 1030~nm. We report, to the best of our knowledge, the highest SH output power generated in the green spectral region using a BiBO crystal with a thin-disk Yb:YAG laser, achieving 32~W at 515~nm from both crystals at a conversion efficiency of 56\%. We systematically characterize the SH output power dependence, long-term power stability, thermal effects on phase matching, SH beam quality, pulse duration, spectral properties, and angular acceptance bandwidth for both materials. Our results provide quantitative, directly comparable performance benchmarks that guide the selection of nonlinear crystals for high-average-power, ultrashort-pulse frequency conversion near 1030~nm.

\section{Experimental setup}

\begin{figure}[t!]
	\centering
	\includegraphics[scale=0.5]{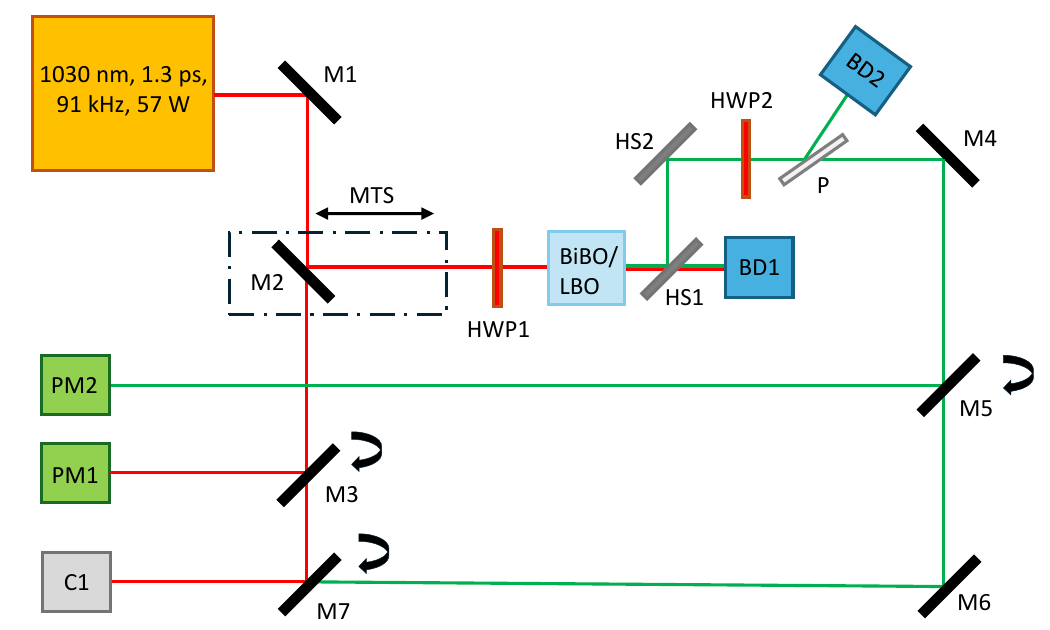}
	\caption{A schematic of experimental setup. M: mirrors; HWP: half waveplates; P: polarizers; HS: harmonic separators (dichroic mirrors); C: camera; PM: power meters; BD: beam dumps; MTS: motorized translation stage}
	\label{fig:Experimental_setup}
\end{figure}

The experimental setup is illustrated in Figure~\ref{fig:Experimental_setup}. The fundamental laser source utilizes a chirped-pulse amplification (CPA) architecture comprising a fiber-based front-end and a Yb:YAG thin-disk regenerative amplifier \cite{app7101016} \cite{Smrz:21}. The system delivers pulses at a central wavelength of 1030~nm with an average output power of 57~W at a repetition rate of 91~kHz and a pulse duration of 1.3~ps. The fundamental laser output beam first passes through an attenuator integrated within the laser housing. This attenuator, comprising a half-wave plate mounted on a motorized rotation stage and two thin-film Brewster polarizers (TFBPs), controls the average power delivered to the downstream optical components on the optical table.

On the optical table, the s-polarized (vertical) component reflected from TFBPs is directed towards the nonlinear crystal via mirrors M1 and M2. Mirror M2 is mounted on a motorized translation stage: when positioned in the beam path, it directs the laser beam to the nonlinear crystal; when retracted, the beam is directed either to power meter PM1 (Ophir F150A-BB-26) via mirror M3 or to beam profiling camera C1 (Cinogy CinCam CMOS 1.001 nano) via mirror M7. Both M3 and M7 are mounted on flip mounts for convenient insertion or removal from the beam path. A half-wave plate HWP is positioned between M2 and the nonlinear crystal to ensure that the fundamental beam enters the crystal with p-polarization (horizontal).

The SH generated in the nonlinear crystal, along with the residual unconverted fundamental, is directed to harmonic separator HS1. At HS1, the SH beam is reflected and directed toward the SH attenuator via harmonic separator HS2, while the transmitted unconverted fundamental is directed to the beam dump BD1. The SH attenuator, consisting of the half-wave plate HWP2, the polarizer P and the beam dump BD2, is used to control the power delivered to the camera and the other diagnostics. The SH beam is then directed either to the power meter PM2 (Ophir 50(150)A-BB-26) via mirrors M4 and M5, or to camera C1 for beam profiling via the M4 and M6 after attenuation. Mirror M5 is also mounted on a flip mount for convenient insertion or removal from the beam path.

The BiBO crystal (Psolutions24) used for SHG has dimensions of [8~mm $\times$ 8~mm] $\times$ 1.5~mm ([aperture] $\times$ length) and is cut at phase-matching angles of $\theta$ = 166.6° and $\phi$ = 90° for type I (e + e $\rightarrow$ o) SHG in the YZ plane. The LBO crystal (Cristal Laser) used for SHG has dimensions of [8~mm $\times$~8 mm] $\times$ 6~mm and is cut at phase-matching angles of $\theta$ = 90° and $\phi$ = 13.5° for type I (o + o $\rightarrow$ e) SHG in the XY plane. The longer interaction length of the LBO crystal (6~mm) was deliberately selected to compensate for its lower effective nonlinear coefficient and thereby yield a SH output comparable to that obtained with the shorter (1.5~mm) BiBO crystal under equivalent pumping conditions. Both entrance and exit faces of the BiBO and LBO crystals are coated with dual-band anti-reflection (DBAR) coatings optimized for 1030~nm and 515~nm. No active temperature stabilization was applied, and all measurements were performed with the crystals maintained at ambient laboratory temperature.

\section{Nonlinear crystal parameters}

\begin{table*}[t!]
	\centering
	\begin{threeparttable}
		\caption{Nonlinear crystal parameters for SHG from 1030~nm to 515~nm at a temperature of 24~°C.}
		\label{BiBO_LBO_comparison}
		\begin{tabular}{l|cc}
			\toprule
			\multicolumn{1}{c}{} & \multicolumn{2}{c}{\textbf{Nonlinear crystals}} \\
			\cmidrule(rl){2-3}
			\textbf{Parameters} & \textbf{BiBO (YZ)} & \textbf{LBO (XY)} \\
			\midrule
			Point group & 2 & mm2 \\
			Crystal length (mm) & 1.5 & 6 \\
			\midrule
			\multicolumn{3}{l}{\textit{Nonlinear and Birefringent Properties}} \\
			\midrule
			Nonlinear coefficient $\textit{d}_\text{eff}$ (pm/V) & 3.4 & 0.828 \\
			$\text{Figure of merit}^{a, b}$ (cm$^2$/W) & $1.3 \times 10^{-9}$ & $1.7 \times 10^{-9}$    \\
			Walk-off angle (mrad) & 30.39 & 8.22 \\
			$\text{Spatial walk-off}^b$ (mm) & 0.046 & 0.049 \\
			$\text{Aperture length}^c$ $l_a$ (mm) & 109 & 402 \\
			\midrule
			\multicolumn{3}{l}{\textit{Group Velocities}} \\
			\midrule
			Group velocity index$^d$ $n_g(\omega)$ @ 1030 nm & 1.824 & 1.626 \\
			Group velocity index$^d$ $n_g(2\omega)$ @ 515 nm  & 1.881 & 1.642 \\			
			Group velocity mismatch$^e$ $\delta v_g^{-1}$ (ps/mm)   & 0.19 & 0.052 \\
			Temporal delay$^b$ (ps)    & 0.29 & 0.31 \\
			$\text{Quasi-static interaction length}^f$ $l_{qs}$ (mm) & 6.8 & 25 \\
			\midrule
			\multicolumn{3}{l}{\textit{Acceptance Bandwidths$^b$ (FWHM)}} \\
			\midrule
			Spectral (nm) & 10.9 & 10.1 \\
			Temperature (K) & 15.8 & 10.5 \\
			Angular (mrad) & 5.6  & 5.8  \\
			\midrule
			\multicolumn{3}{l}{\textit{Absorption}} \\
			\midrule
			$\alpha$ @ 1030 nm (ppm/cm) & 5.1 \cite{Riedel:14}  & 22.9 \cite{Riedel:14} \\
			$\alpha$ @ 515 nm (ppm/cm)  & 312 \cite{Riedel:14}  & 37.3 \cite{Riedel:14} \\
			\bottomrule
		\end{tabular}
		\begin{tablenotes}
			\footnotesize
			\item[$a$] Figure of merit is calculated using 
			$FOM = \frac{{8\pi^2 d_\mathrm{eff}^2 L^2}}{\epsilon_0 \, c \, \lambda_1^2 \, n_\omega^2 \, n_{2\omega}}$,
			where \(L\) is the crystal length, \(n_\omega\) and \(n_{2\omega}\) are the refractive indices at the fundamental and SH frequencies, respectively, \(\lambda_1\) and \(\lambda_2\) denote the fundamental and SH wavelengths, \(c\) is the speed of light in vacuum, and \(\epsilon_0\) is the vacuum permittivity.
			\item[$b$] The figure of merit, spatial walk-off, temporal delay and acceptance bandwidths were calculated using a crystal length of 1.5 mm for BiBO and 6 mm for LBO.		
			\item[$c$] Aperture length defined as $l_a = d / \rho$, where $\rho$ is the walk-off angle and $d = 3.3$ mm is the beam diameter.
			\item[$d$] The group velocity index is obtained by dividing the speed of light by the group velocity. 
			\item[$e$] Group velocity mismatch defined as 
			$\delta v_g^{-1} = [n_g(\omega) - n_g(2\omega)] \, / \, c$ between the fundamental and SH pulses.			
			\item[$f$] Quasi-static interaction length defined as $l_{qs} = \tau_p / \delta v_g^{-1}$, where $\tau_p = 1.3$ ps is the pulse duration.
		\end{tablenotes}
	\end{threeparttable}
\end{table*}

Table~\ref{BiBO_LBO_comparison} summarizes the key crystal and interaction parameters of the BiBO and LBO crystals employed for SHG from 1030~nm to 515~nm. The table includes both intrinsic material properties and derived parameters calculated for the specific experimental conditions used in this work. All the values are based on calculations by mlSNLO software from AS-Photonics \cite{SNLO} unless otherwise stated. The BiBO offers a significantly higher nonlinear coefficient, enabling higher conversion efficiency in shorter crystals. This advantage, however, is accompanied by its large walk-off and strong group velocity mismatch (GVM), which reduce the effective interaction length. Nevertheless, the calculated temporal delays at the crystal exit remain comparable for the chosen crystal lengths, reflecting the shorter BiBO used in this analysis. In contrast, the LBO exhibits much lower walk-off and GVM, resulting in longer spatial and temporal interaction lengths, which allows the use of longer crystals to compensate for its lower nonlinear coefficient. While both crystals show comparable spectral and angular acceptance, the BiBO provides about 50\% wider temperature bandwidth. A key limitation of the BiBO is its high absorption at 515~nm, approximately an order of magnitude higher, leading to an increased thermal load. Consequently, BiBO is advantageous for comparatively compact, high-efficiency conversion, whereas the LBO is the more conservative choice for high-average-power, thermally stable operation.

\section{Experimental results and discussion}

\subsection{Fundamental beam characterization}

\begin{figure}[h!]
	\centering
	\includegraphics[scale=0.35]{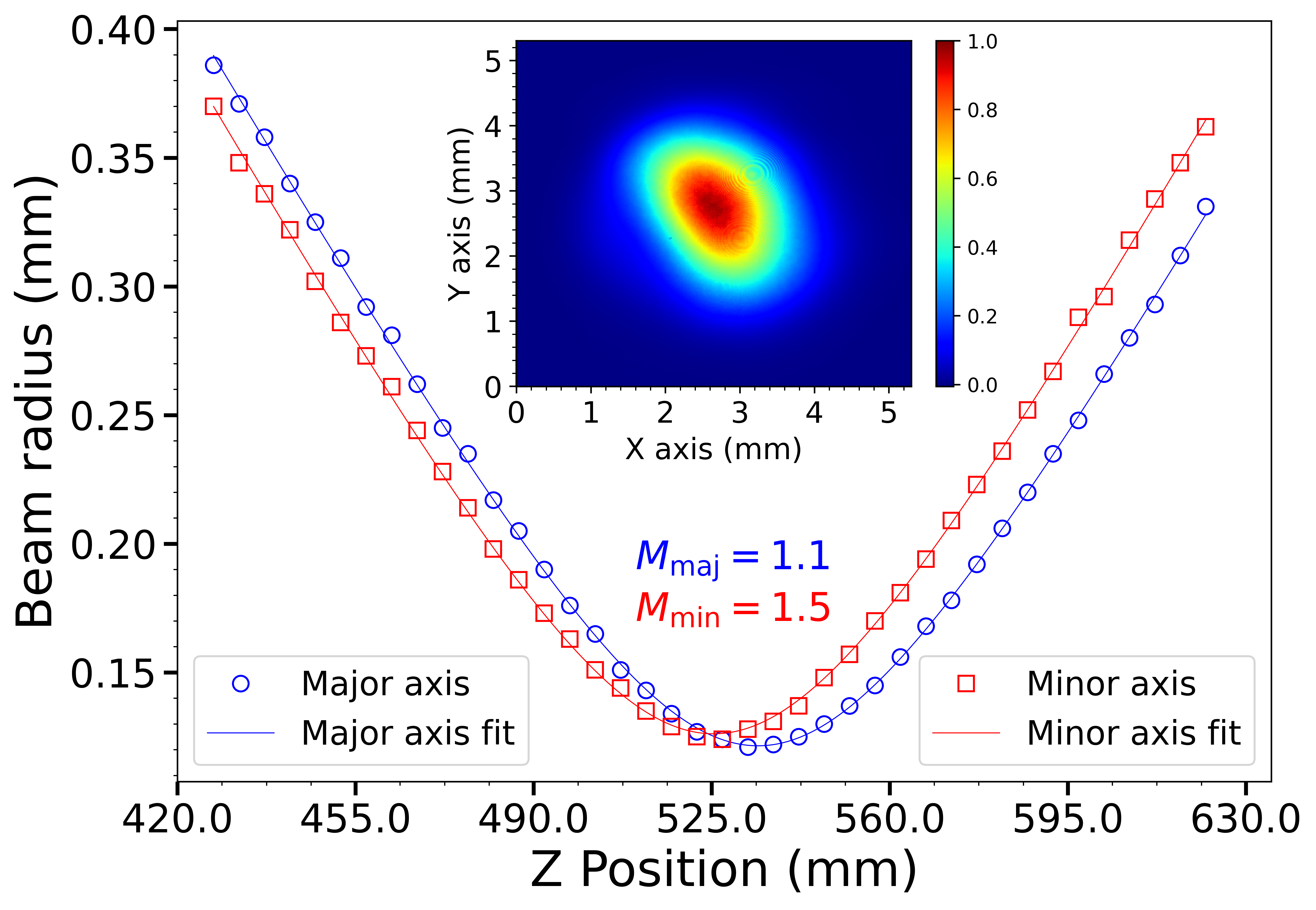}
	\caption{Measured fundamental beam caustic along the major and minor axes, obtained using a focusing lens. Inset shows corresponding near-field beam profile at 57~W.}
	\label{fig:1H_M2_profile}
\end{figure}

The beam quality of the driving fundamental laser for SHG was characterized by measuring the $M^2$ parameter along both principal axes at the beam waist. The measured values were $M^2_{\text{maj}} \! \approx \! 1.1$ and $M^2_{\text{min}} \! \approx \! 1.5$ along the major and minor axes, respectively, indicating a near-diffraction-limited beam with slight astigmatism. The measurement was performed in accordance with the ISO standard for beam quality characterization, and is presented in Figure~\ref{fig:1H_M2_profile}. The spatial profile of the fundamental beam at the crystal input face was also measured, yielding a major axis of 3.3~mm and a minor axis of 2.7~mm, both measured as the second moment beam diameter, corresponding to an elliptical beam shape, as shown in inset of Figure~\ref{fig:1H_M2_profile}. The beam pointing stability was characterized over 25~minutes, demonstrating root mean square (RMS) angular deviations of 11~$\upmu$rad along the x-axis and 10~$\upmu$rad along the y-axis, indicative of excellent pointing stability.

\begin{figure}[h!]
	\centering
	\includegraphics[scale=0.4]{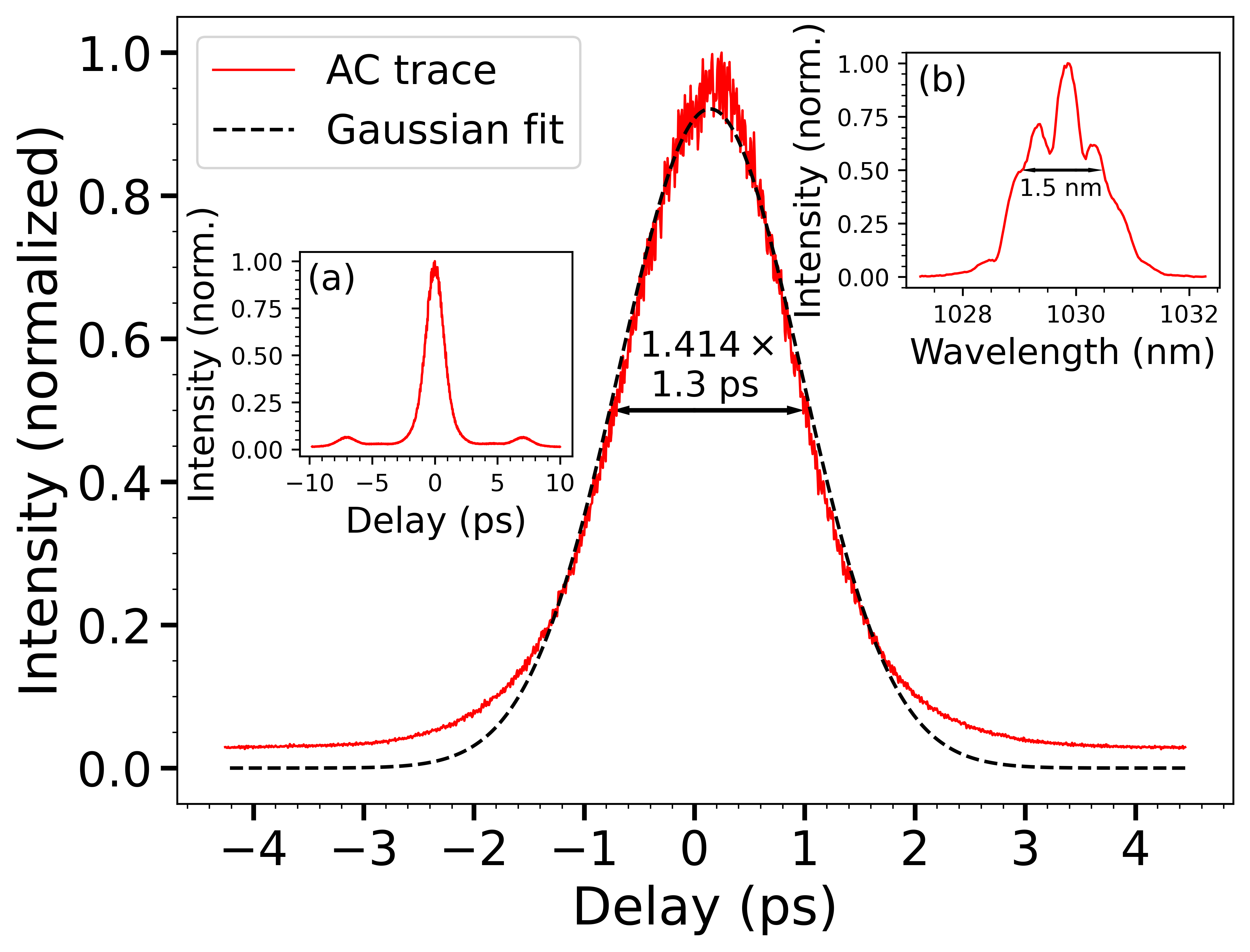}
	\caption{Intensity autocorrelation trace of the amplified 1030 nm laser pulses at the maximum average output power of 57~W. Insets: (a) intensity autocorrelation trace over the full temporal range revealing the presence of side peaks; (b) corresponding spectral distribution of the fundamental centered at 1029.9~nm.}
	\label{fig:1H_AC_spectrum}
\end{figure}

The temporal characteristics of the fundamental pulse were evaluated via intensity autocorrelation measurements (APE, Pulse Check NX50). The measured autocorrelation function (ACF) yielded a FWHM of 1.77~ps. Assuming a Gaussian pulse envelope, the deconvolved pulse duration was determined to be 1.3~ps, as shown in Figure~\ref{fig:1H_AC_spectrum}. The corresponding optical spectrum of the fundamental beam, presented in the inset of Figure~\ref{fig:1H_AC_spectrum}, was acquired using a spectrometer (Narran, BR8) and exhibits a spectral bandwidth of 1.5~nm at FWHM.

\subsection{SH power dependence}

\begin{figure}[h!]
	\centering
	\includegraphics[scale=0.35]{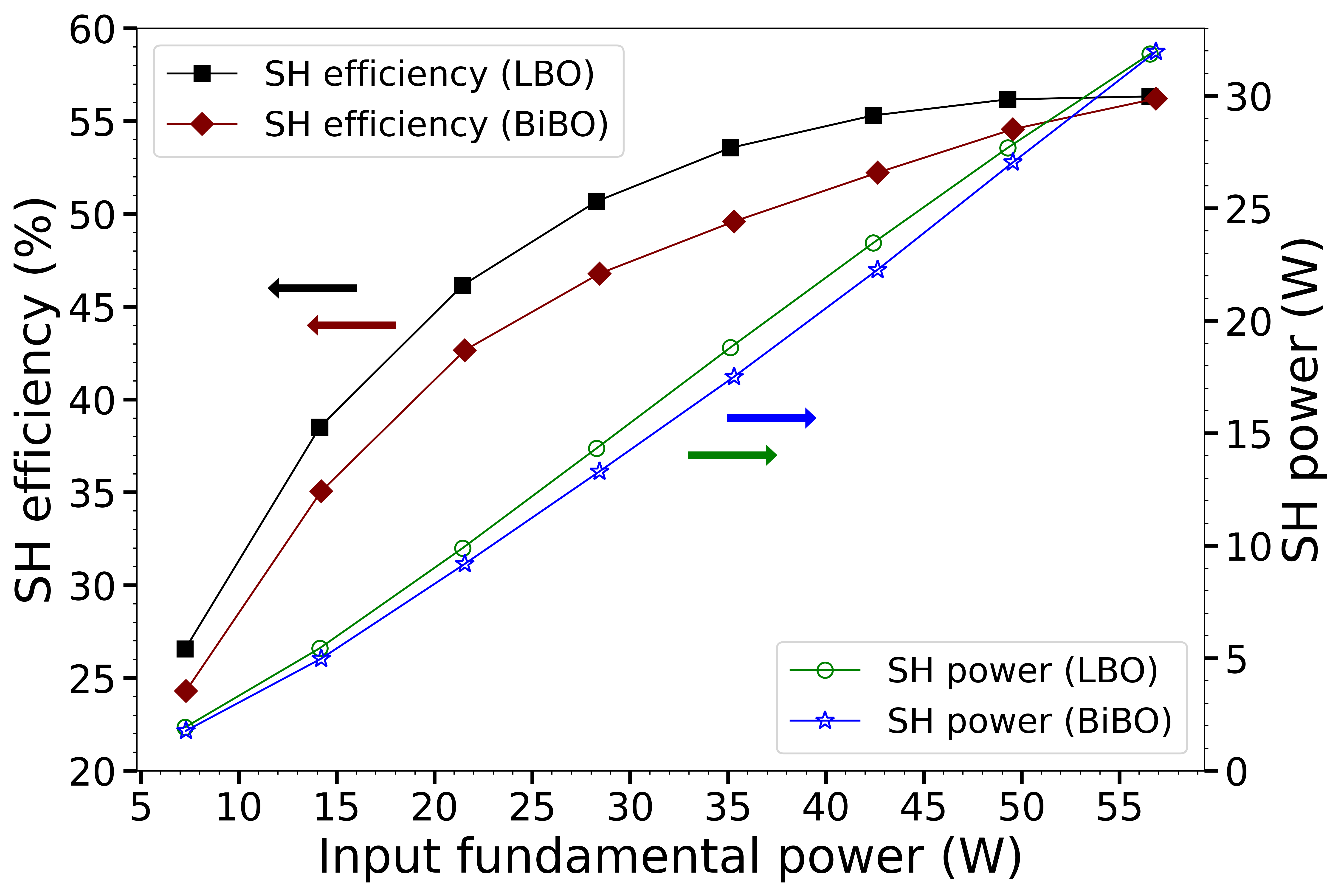}
	\caption{Dependence of SH output power and conversion efficiency on the fundamental input power for SHG in the BiBO and LBO crystals.}
	\label{fig:2H_energy_dependence}
\end{figure}

The SH power dependence was systematically investigated for both the BiBO and LBO nonlinear crystals by incrementally increasing the fundamental input power up to 57~W, corresponding to a peak intensity of $\sim$13~GW/cm$^2$. The measured SH conversion efficiencies and output powers as a function of fundamental input power are presented in Figure~\ref{fig:2H_energy_dependence}. At maximum input power, both crystals yielded 32~W of SH output, corresponding to a conversion efficiency of 56\%. To the best of our knowledge, this represents the highest SH output power generated in the green spectral region using a BiBO crystal and is $\sim$6 times higher than that previously reported by Kumar et al.~\cite{Kumar_2014}. Notably, at all operating points except the highest input power level, both the SH efficiency and output power were lower in the BiBO relative to the LBO. While the SH efficiency in the LBO exhibits clear saturation behavior, the efficiency in the BiBO continues to increase with input power, suggesting that the BiBO had not yet reached its maximum conversion efficiency. Further scaling of the fundamental peak intensity in the BiBO was deliberately avoided due to the comparatively lower optical damage threshold of the crystal. Additionally, the relatively moderate conversion efficiencies observed in both crystals are partly attributed to the presence of side peaks in the temporal profile of the fundamental pulse located approximately 7 ps away from the main pulse, which reduce the effective peak intensity available for the SHG process and thereby limit overall conversion performance.

\subsection{Angular acceptance bandwidth}

\begin{figure}[h!]
	\centering
	\includegraphics[scale=0.37]{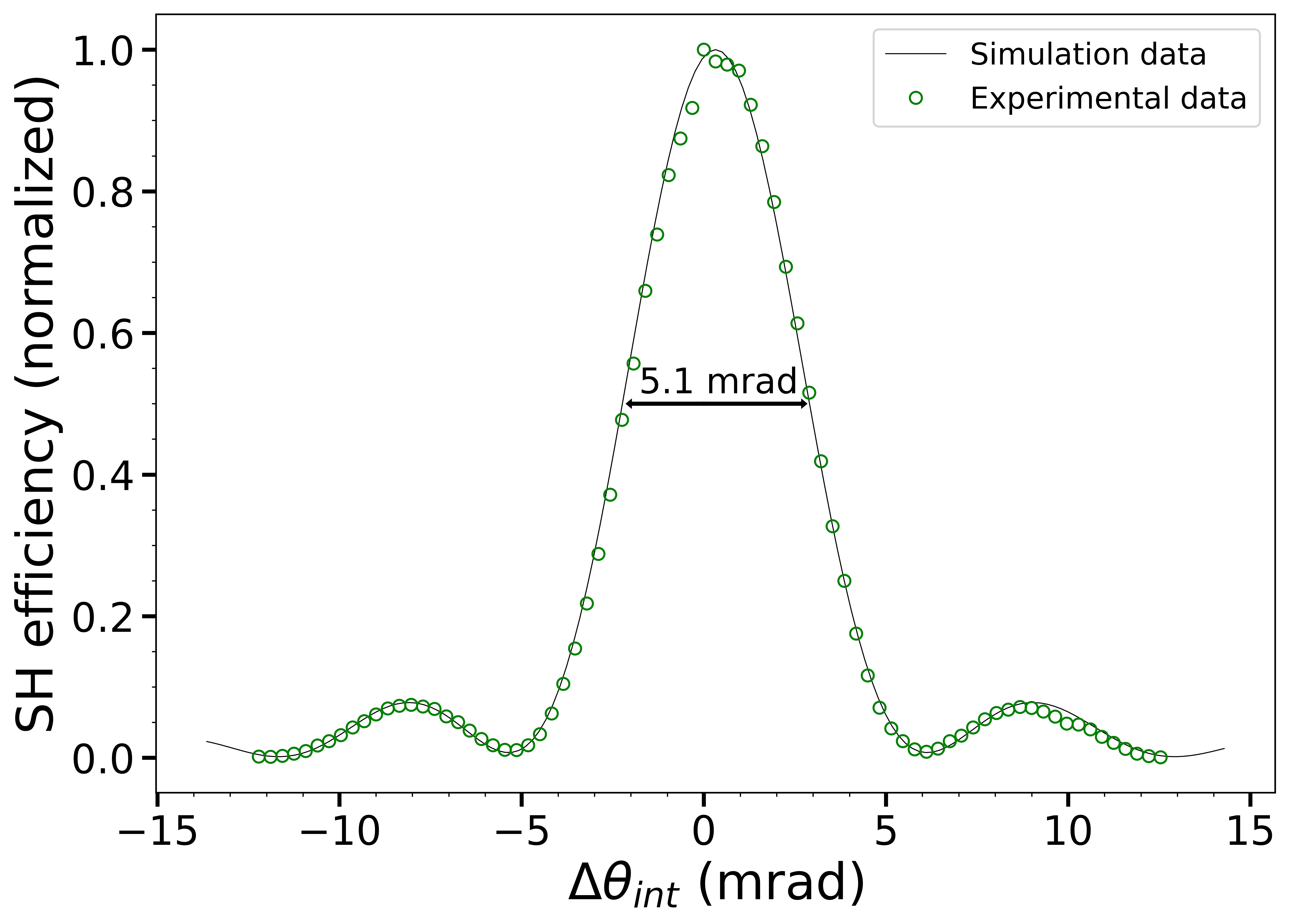}
	\caption{Dependence of the SH conversion efficiency on internal angular detuning for 1.5 mm long BiBO crystal.}	\label{fig:2H_angular_dependene}
\end{figure}

We measured the dependence of the SH power on the rotation of the BiBO crystal about its vertical axis. The fundamental power incident on the crystal was set to 7~W to minimize thermal effects on the measurements, and the SH power was recorded as a function of the phase-matching angle ($\theta$), defined as the angle between the crystal's Z-axis and the internal wave vector of the input beam in the YZ plane. To support the experimental findings, numerical simulations were performed using the mlSNLO software, specifically the 2D-mix-SP module, which accounts for spatial beam profiles and short (picosecond) pulse durations. The simulation methodology follows the modeling approach, in which the nonlinear Maxwell equations are solved using the split-step Fourier method \cite{Smith:95, Smith:1995}, with all simulation inputs derived directly from experimental parameters. The key experimentally derived input parameters used in the simulation were input fundamental power of 7~W, pulse width of 1.3~ps at FWHM, and a beam profile with major and minor axis of 2~mm and 1.6~mm (both at FWHM), respectively. All remaining parameters such as group velocity, or nonlinear coefficient $\textit{d}_\text{eff}$, were taken directly from the values listed in Table~\ref{BiBO_LBO_comparison}. Since the experimentally measured angular changes correspond to external angles, they were converted to internal angles for comparison with the simulated data. The refractive indices of the BiBO crystal at the fundamental and SH wavelengths in the YZ plane, at the angles of interest, were calculated using the Sellmeier equations provided by Miyata et al. \cite{Miyata:09}, and the internal angles were subsequently determined by applying Snell's law. As shown in Figure~\ref{fig:2H_angular_dependene}, the internal angular acceptance bandwidths of the SHG process in the BiBO crystal were found to be approximately 5.1~mrad ($\sim$0.76 mrad$\cdot$cm) and 5.14~mrad ($\sim$0.77 mrad$\cdot$cm) at FWHM for the experimental and simulated data, respectively, indicating excellent agreement between the two. Furthermore, the width of the simulated efficiency curve at the 99\% level corresponds to an angular range of 0.61~mrad, which translates to an external crystal rotation of approximately 1.11~mrad. This implies that the beam pointing fluctuations must remain below $\pm$0.55~mrad to ensure that the SH conversion efficiency does not decrease by more than 1\%. Additional simulations were performed at the maximum fundamental power of 57~W to evaluate the angular acceptance behavior at high conversion efficiencies. The FWHM bandwidth was reduced to 2.69~mrad ($\sim$0.4~mrad$\cdot$cm), nearly half of that obtained at 7~W. This is in good agreement with previously reported observations by Eckardt et al. \cite{1072294}, confirming the pronounced narrowing of the angular acceptance bandwidth at high conversion efficiencies.

An equivalent angular acceptance measurement could not be performed for the LBO crystal because of limited access of the laser. For comparison purposes, simulations were carried out under identical conditions, employing the same input power, pulse duration, and beam diameter as those used in the BiBO simulations. All LBO-specific parameters were taken from Table~\ref{BiBO_LBO_comparison}. The refractive indices of the LBO crystal at both the fundamental and SH wavelengths, evaluated within the XY plane at the relevant phase matching angles, were determined using the Sellmeier equations reported in Kato~\cite{362711}. The simulated internal angular acceptance bandwidth of the SHG process in LBO was found to be approximately 5.25~mrad ($\sim$3.15~ mrad$\cdot$cm) at FWHM. Similar simulations were carried out at the maximum fundamental power of 57~W in order to examine the angular acceptance behavior in the high conversion regime for the LBO crystal. The FWHM bandwidth was found to be 2.3~mrad, indicating a pronounced reduction compared to low power operation. Notably, despite its intrinsically narrower angular acceptance bandwidth per unit length, the 1.5~mm BiBO crystal exhibits a comparable angular acceptance to that of the 6~mm LBO crystal at low power, and an even larger acceptance at full power. This behavior demonstrates that the higher nonlinear coefficient of BiBO compensates for its reduced acceptance per unit length, thereby enabling the use of a significantly shorter crystal without sacrificing angular tolerance.

\subsection{SH power stability}

\begin{figure*}[t!]
	\centering
	\includegraphics[scale=0.043]{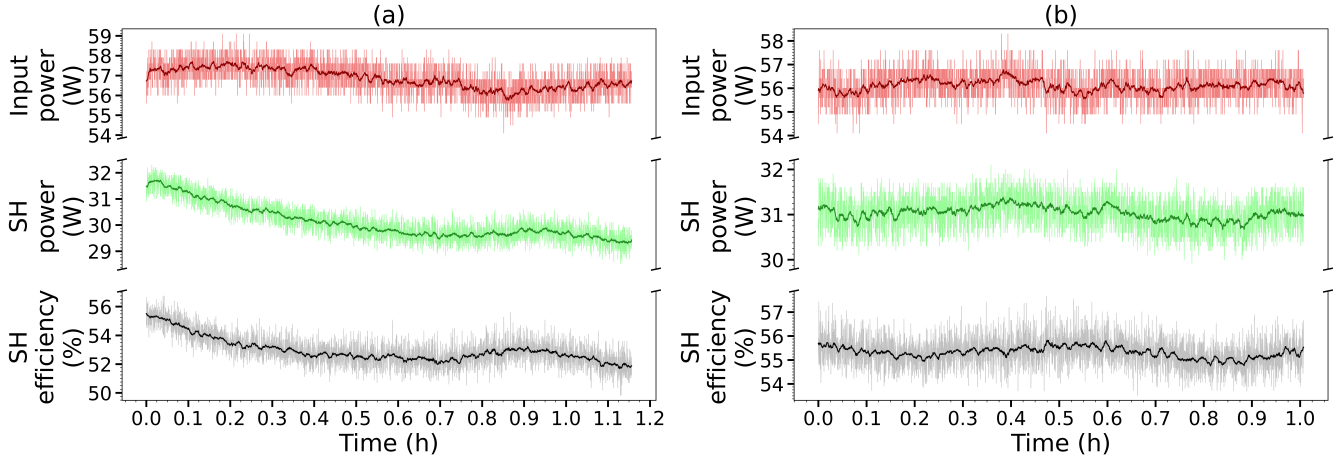}
	\caption{(a) Long-term stability of the SHG in the BiBO crystal measured over 1 hour of continuous operation. (b) Corresponding measurement after re-optimization of the crystal tilt to compensate for thermal effects. In both panels, light colored narrow lines show measured power whereas the darker thick lines show floating average over 500 acquisitions. Note that the vertical scales differ between the two graphs.}
	\label{fig:2H_stability_BiBO}
\end{figure*}

The long-term power stability of the SH output was characterized for both the BiBO and LBO crystals over a duration exceeding one hour. The stability measurement results for the BiBO crystal are presented in Figure~\ref{fig:2H_stability_BiBO}(a), along with a floating average computed over 500 consecutive acquisitions, corresponding to a temporal window of $\sim$30~seconds given the acquisition rate of the power meter, to suppress short-term fluctuations and reveal underlying longer-term trends. Based on this floating-averaged data, a decrease in SH output power of $\sim$2.5~W is clearly observable over the measurement period, corresponding to a reduction of $\sim$8\% in SH output power. This drop is accompanied by a decrease of $\sim$4\% in overall SH conversion efficiency. In contrast, the fundamental power remained comparatively stable within a $\sim$2~W range without any noticeable systematic long-term decrease. This gradual decline is most likely attributable to thermally induced phase mismatch driven by crystal heating, which arises from the relatively high absorption of BiBO at the SH wavelength, as summarized in Table~\ref{BiBO_LBO_comparison}. It should be noted that no active thermal management was employed during these measurements, as the crystal was not housed in a temperature-controlled oven, and it is therefore expected that stabilization of the output power is expected to improve with the implementation of active crystal temperature control. The SH output power stability was quantified as 2.29\% RMS, compared to 1.31\% RMS for the fundamental beam measured over the same period, while the stability of the conversion efficiency was determined to be 1.85\% RMS.

Following this initial characterization, the phase-matching angle ($\theta$) was re-optimized about the vertical axis to recover the original SH output power level, while the laser continued to operate and the SHG process remained active throughout. A second stability measurement was then performed over an additional one-hour period, the results of which are shown in Figure~\ref{fig:2H_stability_BiBO}(b). Since the SHG process had been running continuously for more than one hour prior to the angular re-optimization, the crystal had reached a thermally stable state by the time the second measurement commenced. Consequently, the SH output exhibited significantly improved stability, with the SH output power maintained at $\sim$31~W and a conversion efficiency of $\sim$55\% throughout the entire measurement period, demonstrating excellent long-term stability. The SH output power stability was measured as 1.15\% RMS, the fundamental beam stability as 1.08\% RMS, and the conversion efficiency stability as 0.95\% RMS.

\begin{figure}[h!]
	\centering
	\includegraphics[scale=0.34]{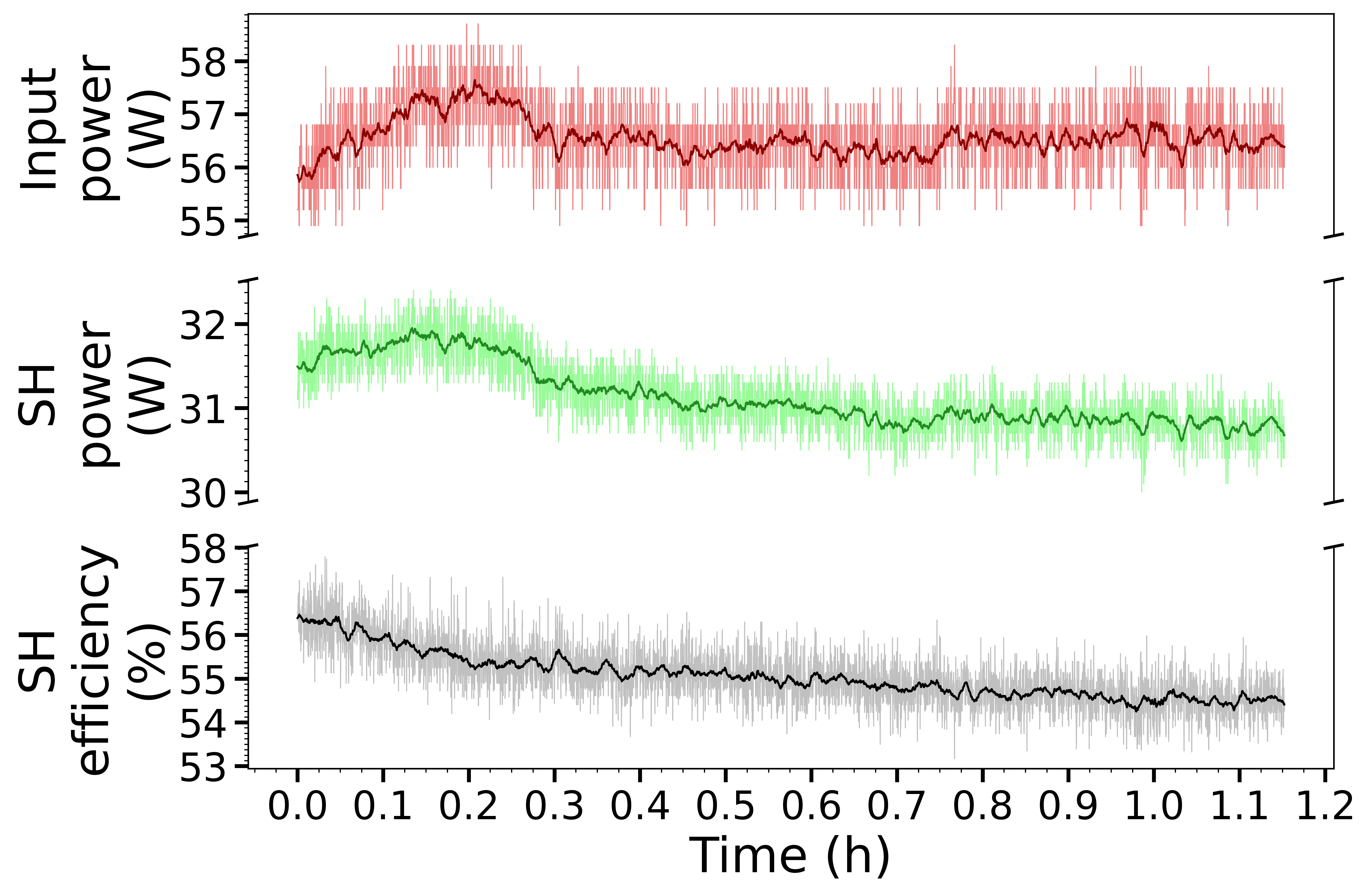}
	\caption{Long-term stability of the SHG in the LBO crystal measured over 1 hour of continuous operation. Light colored narrow lines show measured power whereas the darker thick lines show floating average over 500 acquisitions.}	
	\label{fig:2H_stability_LBO}
\end{figure}

In comparison, the LBO crystal exhibited superior SH output power stability, as presented in Figure~\ref{fig:2H_stability_LBO}. Based on floating-averaged data, a decrease in SH output power of $\sim$0.8 W was also observed in the LBO over the measurement period, corresponding to a reduction of $\sim$2.5\% in SH output power. The associated drop in overall SH conversion efficiency was of $\sim$2\%, while the fundamental power remained comparatively stable within a $\sim$2~W range without any noticeable long-term decrease. However, the magnitude of this drift was smaller than that recorded for the BiBO, reflecting the improved thermal behavior of LBO. This is probably due to the lower absorption of LBO at the SH wavelength relative to that of BiBO, as summarized in Table~\ref{BiBO_LBO_comparison}, resulting in reduced crystal heating and consequently more stable phase-matching conditions. The SH output power stability for the LBO was measured as 1.34\% RMS, with the fundamental beam exhibiting 1.09\% RMS stability over the same period, and the conversion efficiency stability quantified as 1.14\% RMS.

\subsection{Thermal effects on phase matching and SH power stability in BiBO}

\begin{figure}[h!]
	\centering
	\includegraphics[scale=0.37]{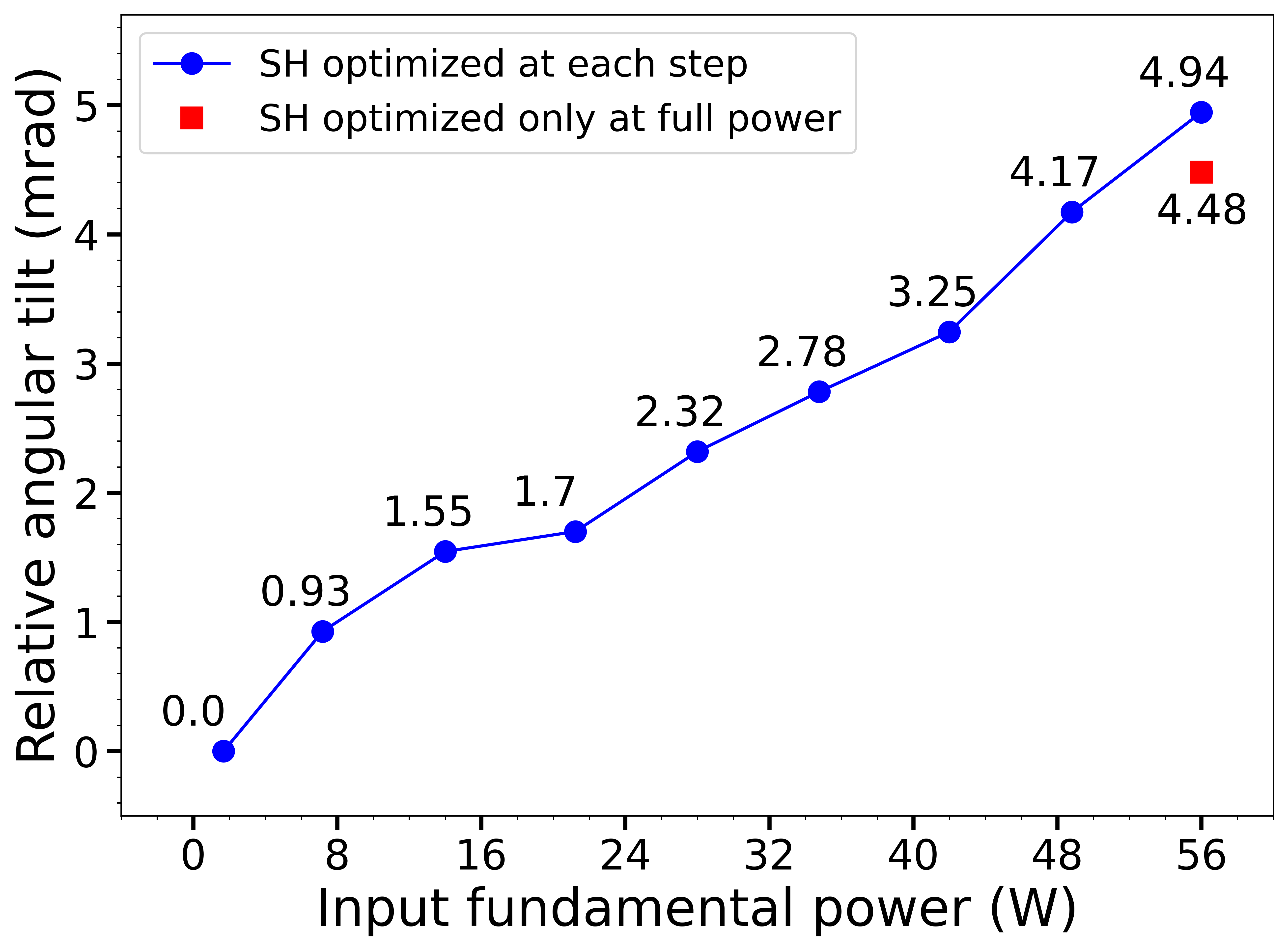}
	\caption{Relative external angular tilt required for SH optimization as a function of input fundamental power. Blue circles show the tilt needed when the SH is re-optimized at each power step. The red square indicates the tilt when optimized only at full power.}	
	\label{fig:Thermal_effects_BiBO}
\end{figure}

To investigate thermal effects on SH power in a BiBO crystal arising from absorption during the SHG process, we systematically characterized the shift in phase-matching angle that provides the highest SH output power as a function of incident fundamental power, as illustrated in Figure~\ref{fig:Thermal_effects_BiBO}. The crystal rotation was initially optimized for phase matching at minimal fundamental input power ($\sim$1.5~W) to establish a thermally unperturbed reference condition, defined as zero angular displacement. Upon increasing the input to full fundamental power and re-optimizing the phase-matching condition, an external angular correction of $\sim$4.5~mrad was required, corresponding to an internal angular shift of $\sim$2.5~mrad. This measurable angular displacement indicates thermally induced changes in the refractive indices. The heating of the crystal is due to the absorption of the beams in the bulk and in the AR coatings \cite{Stubenvoll_2016, Riedel:14}.

To further characterize the thermal dynamics, the crystal was returned to the reference position and allowed to fully thermalize to ambient conditions. The fundamental power was then increased incrementally, with the phase-matching angle re-optimized at each power step relative to the reference. Upon reaching full power through this stepwise procedure, the total external angular displacement reached $\sim$5~mrad, corresponding to an internal shift of $\sim$2.7~mrad, which represents a discrepancy of $\sim$0.5~mrad (external) relative to the direct ramp measurement. This difference arises from two contributing factors. First, the extended interaction time afforded by the incremental approach, during which cumulative heating of the crystal resulted in a greater degree of thermalization prior to reaching the final power level. In contrast, the rapid transition to full power in the direct measurement did not allow sufficient time for the crystal to reach a thermally stable state, resulting in a smaller observed angular shift at the moment of measurement. Second, the width of internal angular acceptance bandwidth at the 99\% SH efficiency level spans only $\sim$1.11~mrad ($\pm$0.55~mrad), meaning that small deviations from the true phase-matching angle produce negligible changes in SH output power within this narrow acceptance window; i.e., both values of maximum power are within error of the measurements. As a result, resolving angular misalignments from ideal phase matching with sub-mrad precision through power optimization alone is inherently difficult.

\begin{figure}[h!]
	\centering
	\includegraphics[scale=0.34]{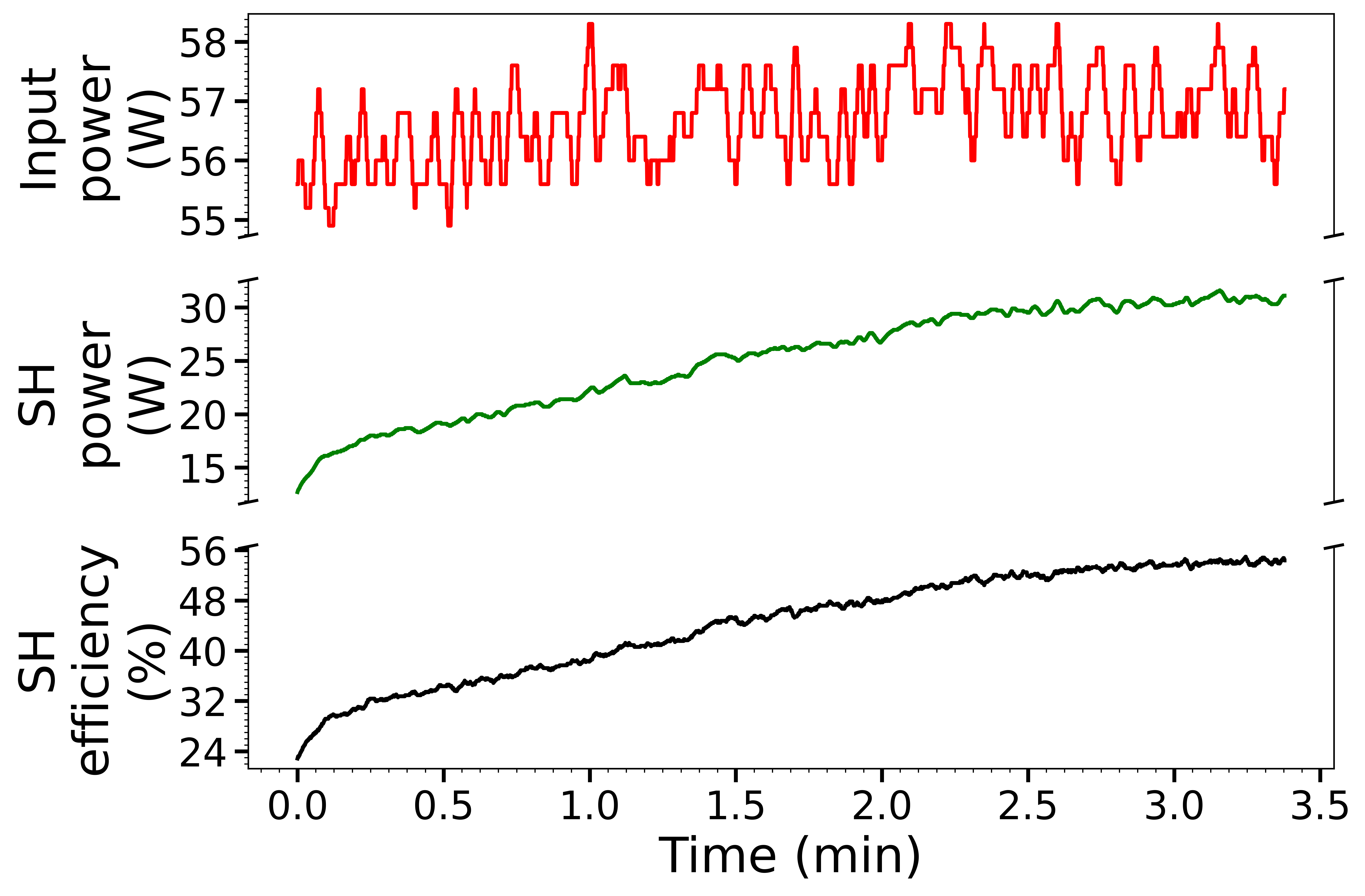}
	\caption{Temporal evolution of the SHG process during thermalization of the BiBO crystal immediately after the onset of SHG. The phase-matching angle was optimized in the previous run after the crystal had thermally stabilized at full power. In the present measurement, the crystal temperature was initially equal to the ambient temperature.}	
	\label{fig:2H_evolution}
\end{figure}

Following this, the fundamental beam was redirected to the beam dump within the attenuator to allow the nonlinear crystal to cool and re-equilibrate to the ambient laboratory temperature. The following measurement was performed to reconfirm the presence of thermally induced effects within the crystal. After crystal cooling and thermal stabilization, the full fundamental power was quickly reintroduced into the crystal to evaluate the recovery of the SH output, with the result presented in Figure~\ref{fig:2H_evolution}. At the onset, the SH power was lower as the crystal temperature had returned to ambient and was no longer aligned with the phase-matching condition optimized for elevated operating temperatures. However, as the SHG process continued, the absorption of the generated SH led to gradual heating of the crystal. This thermally induced shift in the refractive indices brought the crystal to the optimal phase-matching condition. Consequently, a progressive increase in both SH power and conversion efficiency was observed over time, ultimately reaching the previous performance levels, thereby confirming the presence of thermal effects in the crystal.

To establish a quantitative correspondence between the observed angular shifts and the underlying thermal load, the measured internal phase-matching angle shift of $\sim$2.5~mrad was converted to an equivalent crystal temperature rise using the known temperature-dependent Sellmeier equations for BiBO. This analysis yields an estimated temperature increase of $\sim$7~$^{\circ}$C relative to the ambient reference condition. To independently validate this estimate, the crystal surface temperature was monitored using an infrared (IR) thermal imaging camera (FLIR E50). Prior to the experiment the crystal temperature was recorded at $\sim$20~$^{\circ}$C, consistent with the laboratory ambient. Upon optimization of the phase-matching angle at full fundamental power and after the crystal had reached a thermally stable state, the IR camera recorded a surface temperature of $\sim$27~$^{\circ}$C, corresponding to a net temperature rise of $\sim$7~$^{\circ}$C. The agreement between the thermally inferred angular shift and the directly measured surface temperature increment provides independent confirmation that the observed phase-matching angle drift is predominantly due to heating of the crystal, and that the crystal temperature rise under full operating power is well characterized by this $\sim$7~$^{\circ}$C increase.

\subsection{SH beam characterization}

\begin{figure*}[h!]
	\centering
	\includegraphics[scale=0.045]{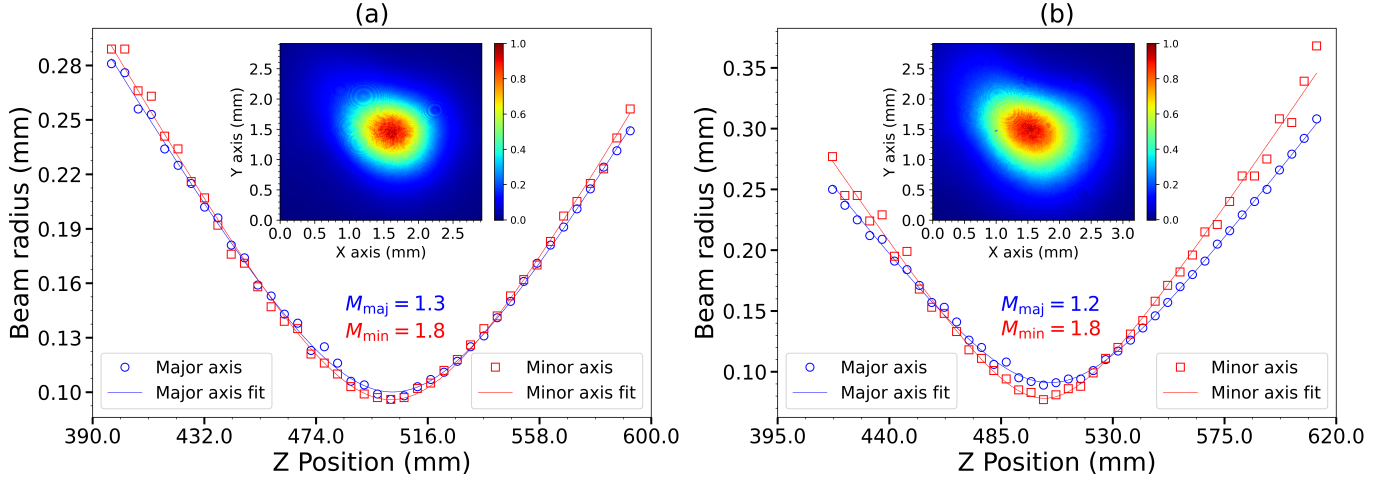}
	\caption{(a) Measured SH beam caustic along the major and minor axes generated in the BiBO crystal, obtained using a focusing lens. Inset shows the corresponding near-field beam profile of the SH output measured at 32~W. (b) Measured SH beam caustic along the major and minor axes generated in the LBO crystal, obtained using a focusing lens. Inset shows the corresponding near-field beam profile of the SH output measured at 32~W.}
	\label{fig:BiBO+LBO_M2_beam_profiles}
\end{figure*}

The beam quality factor $M^2$ of the SH beam, generated at the highest fundamental input power by both the BiBO and LBO crystals was characterized following the same procedure as for the fundamental beam. For the BiBO crystal, the measured beam quality factors were $M^2_{\text{maj}} \! \approx \! 1.3$ and $M^2_{\text{min}} \! \approx \! 1.8$ along the major and minor axes, respectively, while for the LBO crystal, values of $M^2_{\text{maj}} \! \approx \! 1.2$ and $M^2_{\text{min}} \! \approx \! 1.8$ along the major and minor axes, respectively, were obtained. The caustic measurements of the SH beam generated in the BiBO and LBO crystals are presented in Figures~\ref{fig:BiBO+LBO_M2_beam_profiles}(a) and \ref{fig:BiBO+LBO_M2_beam_profiles}(b), respectively, with the corresponding near field spatial intensity profiles shown as insets. The beam quality characterization was performed in accordance with the ISO standard for $M^2$ measurements. The degradation in beam quality of the SH relative to the fundamental beam is primarily attributed to spatial imperfections in the fundamental beam profile. Since the SHG process directly depends on the fundamental intensity and its phase, any spatial imperfections in the fundamental beam affect SHG, resulting in increased beam distortion and higher $M^2$ values.

The SH near-field spatial profile of BiBO is smaller compared to LBO. The measured beam widths, evaluated as second moment beam diameter, are 2.6~mm (major axis) and 2.1~mm (minor axis) for BiBO, whereas LBO exhibits 3.1~mm (major axis) and 2.3~mm (minor axis), respectively. This corresponds to an approximately $\sim$25\% reduction in beam area for BiBO. This smaller spatial profile in BiBO can be explained by thermally induced phase-matching variations caused by the Gaussian intensity distribution of the fundamental beam. The crystal is optimized for phase matching near the beam center, where the intensity and thermal load are maximum. However, the resulting radial temperature gradient leads to local changes in phase mismatch, which reduces SH conversion toward the beam edges, resulting in the observed reduction in the SH beam size.

\begin{figure*}[h!]
	\centering
	\includegraphics[scale=0.05]{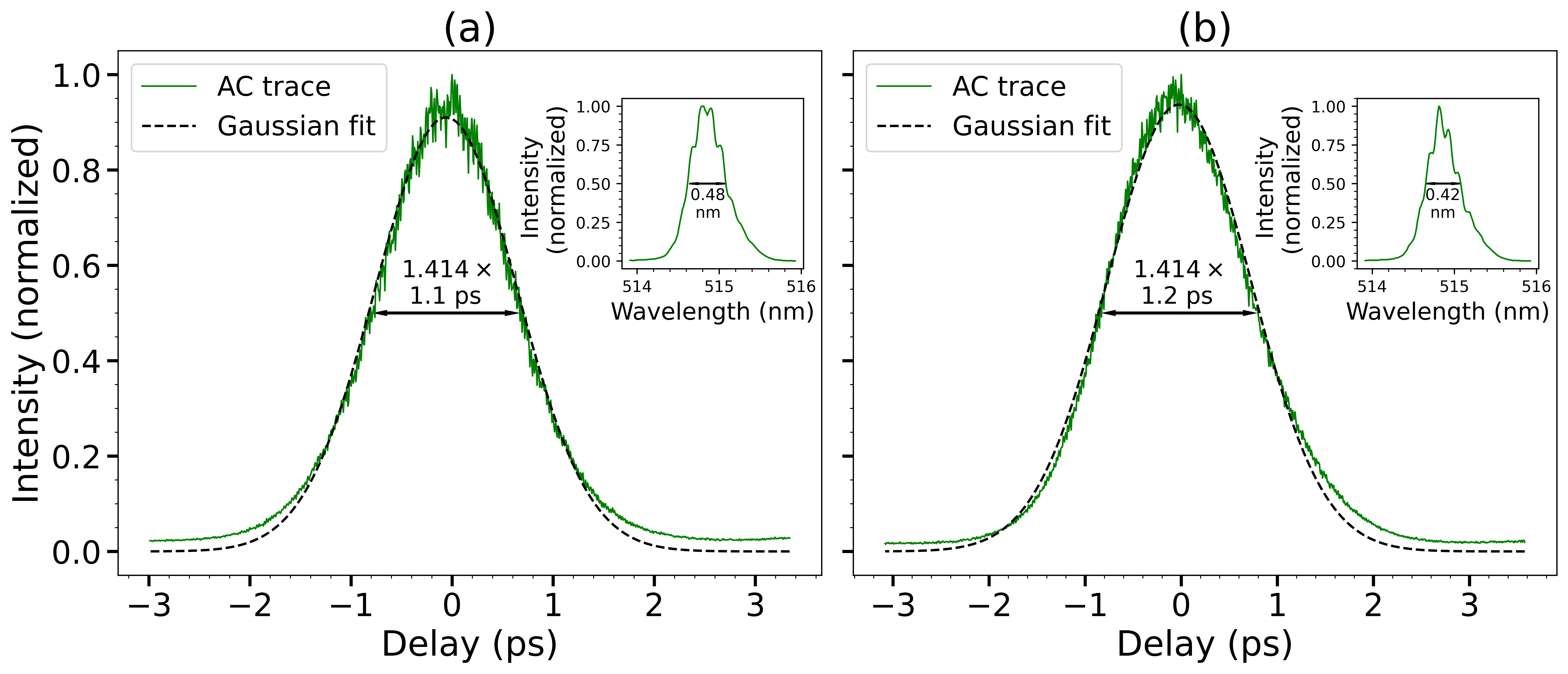}
	\caption{(a) Intensity autocorrelation trace of the SH pulses generated in the BiBO crystal at an SH output power of 32~W. Inset: corresponding spectral distribution of the SH centered at 514.8~nm. (b) Intensity autocorrelation trace of the SH pulses generated in the LBO crystal at an SH output power of 32~W. Inset: corresponding spectral distribution of the SH centered at 514.8~nm.}
	\label{fig:2H_AC_spectrum}
\end{figure*}

To characterize the SH pulses generated in the BiBO and LBO crystals, both autocorrelation traces and spectra were measured. The autocorrelation traces and corresponding spectral distributions (shown as insets) for the BiBO and LBO crystals are presented in Figures~\ref{fig:2H_AC_spectrum}(a) and (b), respectively. The measured ACF widths (FWHM) of the SH pulses are 1.55~ps and 1.67~ps for the BiBO and LBO, respectively. Assuming a Gaussian pulse envelope, the deconvolved pulse durations (FWHM) are determined to be 1.1~ps for the BiBO and 1.2~ps for the LBO, with the shorter duration obtained from the BiBO primarily attributed to the difference in crystal length required for equivalent conversion efficiency. Despite its larger GVM per unit length, BiBO's higher effective nonlinear coefficient ($\textit{d}_\text{eff}$) enables efficient SHG in a shorter crystal. Consequently, the longer LBO crystal leads to greater temporal walk-off accumulation between the fundamental and SH pulses due to GVM, amounting to an additional 0.02~ps of temporal separation, which contributes to a longer SH pulse duration.

The SH spectrum in the BiBO (0.48~nm) is $\sim$15\% broader than in the LBO (0.42~nm). This is due to its broader spectral acceptance bandwidth, caused by reduced GVM between the fundamental and SH pulses (see Table~\ref{BiBO_LBO_comparison}). Since the fundamental pulse carries a finite spectral bandwidth, BiBO's intrinsic material properties allow it to accept and efficiently convert a larger portion of that spectrum into the SH, as evidenced by its ~7\% broader spectral acceptance bandwidth of compared to LBO (see Table~\ref{BiBO_LBO_comparison}). This broader spectral acceptance simultaneously explains the slightly shorter pulse duration observed in BiBO, with the pulse duration being ~0.1~ps shorter, as wider spectral bandwidth in the frequency domain corresponds to a shorter pulse in the time domain.

\section{Conclusion}

In summary, we have demonstrated and directly compared SHG in the 1.5~mm long BiBO and 6~mm long LBO crystals under identical high-average-power operating conditions, achieving 32 W of green output at 515~nm with a conversion efficiency of 56\% from both crystals, where the higher effective nonlinearity of BiBO enabled equally efficient conversion in a significantly shorter crystal length compared to LBO. The SH pulse duration from the BiBO was measured to be 1.1~ps with a spectral bandwidth of 0.48~nm, slightly shorter and broader than the 1.2~ps pulses with 0.42~nm bandwidth obtained from the LBO, which is attributed to BiBO's larger spectral bandwidth and reduced GVM accumulation. Despite its higher effective nonlinearity and broader angular acceptance bandwidth, BiBO showed greater susceptibility to thermally induced phase-mismatch drift, requiring angular re-optimization upon changes in incident power. LBO, by virtue of its lower absorption at the SH wavelength, proved to be the more thermally stable option, exhibiting less output power drift over extended operation near a wavelength of 1030 nm. The experimentally measured and numerically simulated angular acceptance bandwidths of SHG in BiBO show excellent agreement, confirming the reliability of the measurement. Despite its shorter length, the BiBO crystal exhibits an angular acceptance comparable to that of the longer LBO crystal. For applications requiring long-term, hands-off operation, housing the BiBO crystal in a temperature-controlled oven is recommended to suppress thermally driven phase-mismatch variations and achieve improved stability. Overall, both crystals represent viable solutions for efficient green light generation from 100 W-class picosecond lasers near wavelength of 1030~nm, with the optimal choice depending on the specific requirements of the target application.

\begin{center}\textbf{Acknowledgement}\end{center}

\noindent This work was supported by the European Union and the state budget of the Czech Republic under the project LasApp \nolinebreak (CZ.02.01.01/00/22\textunderscore008/0004573). The authors would like thank Yuya Koshiba for valuable discussions and careful proofreading of the manuscript.

\bibliography{sample}

\end{document}